# Modal synchronization of coupled bistable Van der Pol oscillators.

I.B. Shiroky and O.V. Gendelman*

Faculty of Mechanical Engineering, Technion – Israel Institute of Technology,

Haifa, 3200003, Israel

* - contacting author, ovgend@technion.ac.il

The paper revisits recently revealed regimes of the "nonconventional synchronization" in systems of coupled bi-stable Van der Pol oscillators. These regimes are characterized by periodic (or quasiperiodic) almost complete energy exchanges between the coupled oscillators. In the paper it is demonstrated that such responses correspond to synchronization of the modulation amplitudes between symmetric and antisymmetric modes of the system, with persistent phase drift. This observation substantially simplifies the treatment, reduces the dynamics to simple phase cylinder and allows to reveal a rich set of global and local bifurcations of limit cycles and tori in the system. Among other findings, one encounters an unexpected regime of nonstationary asymmetric synchronization.

1. Introduction

Van der Pol (VDP) oscillator entered scientific literature in 1920 [1], and became a paradigmatic model for systems with the limit cycle (LC) as a single attractor. In the case of biophysics, the model has been used to study heart beats [2], neurobiology of the lamprey [3], Parkinsonian tremor [4] , eye chemistry [5, 6], EEG dynamics [7], vocal fold oscillator during phonation [8, 9], action potentials of neurons [10] and others. Multiple additional applications of the model include as different subjects as radio electronics, seismology, nonlinear optics and many others. Modified versions of the VDP oscillator can include higher orders of the nonlinear damping. In particular, it is easy to formulate a bi-stable variation of the VDP oscillator with two attractors: a stable fixed point and a stable limit cycle separated by an unstable limit cycle [11]. Discrete chains of bi-stable van der Pol oscillators were studied in [12, 11] where basic regimes of localization and front propagation were demonstrated. .

Phenomenon of synchronization in systems of coupled oscillators is well-known in scientific literature [13, 3, 14, 15, 16]. In most cases, one observes a synchronization of stationary responses of the oscillators. As for the coupled VDP oscillators, studies [3, 14] and many others revealed the synchronization regimes when the excitation amplitude of each oscillator remained constant. The case of strongly coupled VDP oscillators is addressed in [17] and compared to the weakly coupled case. This type of synchronization is classified in resent works as being close to nonlinear normal modes [15, 18]. Recently, lot of attention has been paid to nonstationary regimes with strong energy exchange in coupled oscillatory systems [19, 20, 21]. In particular, recently it was revealed [15, 16, 22] that the system of linearly coupled bistable VDP oscillators exhibits quite unusual pattern of nonstationary "non-conventional" synchronization. In this regime the response of the oscillators is strongly modulated; one can say that they exchange excitation. This regime cannot be described in terms of classical synchronization that primarily deals with the stationary responses. The analytical approach used in [15] was based on special symmetry revealed by the system of slow-flow equations for one special combination of the system parameters. In this case, the slow-flow system was reduced to the flow on the phase plane, and the regime of the "non-conventional" synchronization has been clearly identified as the limit cycle of the slow flow. Further numerical studies demonstrated that this regime robustly appeared for finite region in the parametric space, and not only for the special combination of parameters that endowed the slow-flow equations with the special symmetry. Paper [16] explored the effect of detuning in a similar model. In [22] additional effects of the damping and nonlinearity were addressed. The authors showed that for same sets of parameters two regimes can co-exist – the stationary synchronous dynamics and the "non-conventional" synchronization.

Current paper offers an alternative approach to the same model of two weakly coupled bi-stable VDP oscillators. It will be demonstrated that that these special regimes of the "non-conventional" synchronization correspond to synchronization of the modulation amplitudes between modal coordinates of the system, with persistent phase drift. We are going to demonstrate that this observation drastically simplifies the analysis and allows exhaustive parametric study of the system without reliance on any nontrivial and parameter-dependent symmetry. Then, it is possible to reveal rich bifurcation patterns, as well as the regions in the space of parameters where each of the qualitatively different synchronization regimes is stable. The structure of the paper is as follows. In section 2 the model is described, and the regime of "non-conventional" synchronization is presented in physical variables, i.e. in terms of displacements of the individual oscillators. In section 3 the system is analyzed in the modal space and different response regimes and their bifurcations are described. In section 4 the results are verified numerically versus the original model. Section 5 is devoted to conclusions and discussion.

2. **Analysis of the model in physical coordinates**

We consider a model of two linearly coupled identical bi-stable VDP oscillators similar to models previously discussed in [11, 12, 15, 16, 22]:

$$y_1'' + y_1 + \varepsilon c(y_1 - y_2) + \varepsilon y_1'(1 - y_1^2 + \eta y_1^4) = 0$$
$$y_2'' + y_2 + \varepsilon c(y_2 - y_1) + \varepsilon y_2'(1 - y_2^2 + \eta y_2^4) = 0$$
(1)

Here $c$ is the coupling between the oscillators, mass, linear frequency and the coefficient of quadratic term in the nonlinear damping are set to unity without loss of generality. Both the coupling and the nonlinear damping are assumed as small, and $0 < \varepsilon \ll 1$ is the appropriate parameter. First, we numerically demonstrate the regime of strong modulation of the fast oscillations with almost complete energy exchange between the two oscillators (Figure 1). The system was deliberately simulated for a large value of $\varepsilon = 0.05$ in order to be able to identify visually the fast oscillations in the plot.

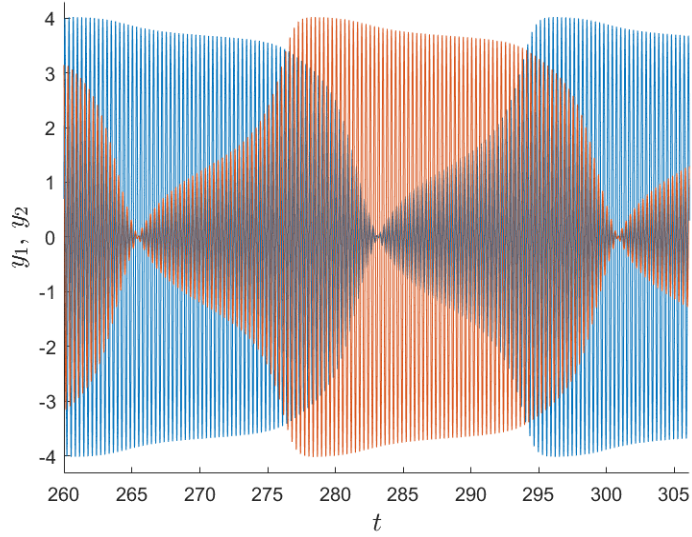

*Figure 1 – The regime of "non-conventional" synchronization, $y_1$ - blue, $y_2$ - red (time rescaled by 20), Parameters: $c = 0.3$, $\eta = 0.1$, $\varepsilon = 0.05$.*

The following slow-flow analysis is similar to the one used in [15, 16, 22] and many other works, and is presented here for the sake of completeness. In order to separate the slow evolution of the envelope from the fast oscillations through common analytical techniques [19, 20, 21], complex variables $\Theta_1$, $\Theta_2$ are introduced:

$$y_1 = -i\frac{\Theta_1 - \Theta_1^*}{2}, \quad y_1' = \frac{\Theta_1 + \Theta_1^*}{2}, \quad y_1'' = \Theta_1' - i\frac{\Theta_1 + \Theta_1^*}{2}$$
$$y_2 = -i\frac{\Theta_2 - \Theta_2^*}{2}, \quad y_2' = \frac{\Theta_2 + \Theta_2^*}{2}, \quad y_2'' = \Theta_2' - i\frac{\Theta_2 + \Theta_2^*}{2} \qquad (2)$$

Substitution of (2) into (1) yields:

$$\Theta_1' - i\Theta_1 - \frac{i\varepsilon c}{2}\left(\Theta_1 - \Theta_2 - \Theta_1^* + \Theta_2^*\right) + \varepsilon\left(\frac{\Theta_1 + \Theta_1^*}{2}\right)\left(1 + \left(\frac{\Theta_1 - \Theta_1^*}{2}\right)^2 + \eta\left(\frac{\Theta_1 - \Theta_1^*}{2}\right)^4\right) = 0$$

$$\Theta_2' - i\Theta_2 + \frac{i\varepsilon c}{2}\left(\Theta_1 - \Theta_2 - \Theta_1^* + \Theta_2^*\right) + \varepsilon\left(\frac{\Theta_2 + \Theta_2^*}{2}\right)\left(1 + \left(\frac{\Theta_2 - \Theta_2^*}{2}\right)^2 + \eta\left(\frac{\Theta_2 - \Theta_2^*}{2}\right)^4\right) = 0 \qquad (3)$$

By substitution of $\Theta_1 = \phi_1 e^{it}$, $\Theta_2 = \phi_2 e^{it}$ into (3) and by dividing the equation by $e^{i\omega t}$ we achieve:

$$\phi_1' - \frac{i\varepsilon c}{2}\left(\phi_1 - \phi_2 - \phi_1^* e^{-2it} + \phi_2^* e^{-2it}\right) +$$

$$+\varepsilon \frac{\phi_1 + \phi_1^* e^{-2it}}{2}\left(1 + \frac{\phi_1^2 e^{2it} - 2|\phi_1|^2 - (\phi_1^*)^2 e^{-2it}}{4} + \eta \frac{\phi_1^4 e^{4it} - 4|\phi_1|^2 \phi_1^2 e^{2it} + 6|\phi_1|^4 - 4|\phi_1|^2 (\phi_1^*)^2 e^{-2it} + (\phi_1^*)^4 e^{-4it}}{16}\right) = 0$$

$$\phi_2' + \frac{i\varepsilon c}{2}\left(\phi_1 - \phi_2 - \phi_1^* e^{-2it} + \phi_2^* e^{-2it}\right) +$$

$$+\varepsilon \frac{\phi_2 + \phi_2^* e^{-2it}}{2}\left(1 + \frac{\phi_2^2 e^{2it} - 2|\phi_2|^2 - (\phi_2^*)^2 e^{-2it}}{4} + \eta \frac{\phi_2^4 e^{4it} - 4|\phi_2|^2 \phi_2^2 e^{2it} + 6|\phi_2|^4 - 4|\phi_2|^2 (\phi_2^*)^2 e^{-2it} + (\phi_2^*)^4 e^{-4it}}{16}\right) = 0$$

(4)

The fast oscillations are averaged out from (4) [23, 12], and one obtains the following set of equations for the slow flow:

$$\phi_1' - \frac{i\varepsilon c}{2}(\phi_1 - \phi_2) + \frac{\varepsilon \phi_1}{2}\left(1 - \frac{|\phi_1|^2}{4} + \eta \frac{|\phi_1|^4}{8}\right) = 0$$

$$\phi_2' - \frac{i\varepsilon c}{2}(\phi_2 - \phi_1) + \frac{\varepsilon \phi_2}{2}\left(1 - \frac{|\phi_2|^2}{4} + \eta \frac{|\phi_2|^4}{8}\right) = 0$$

(5)

Let us define a new time scale such that:

$$\tau = \varepsilon t \rightarrow \phi' = \frac{d\phi}{dt} = \varepsilon \frac{d\phi}{d\tau} = \varepsilon \dot\phi \tag{6}$$

Substitution of (6) into (5) leads to elimination of $\varepsilon$ from the equations:

$$\dot\phi_1 - \frac{ic}{2}(\phi_1 - \phi_2) + \frac{\phi_1}{2}\left(1 - \frac{|\phi_1|^2}{4} + \eta \frac{|\phi_1|^4}{8}\right) = 0$$

$$\dot\phi_2 - \frac{ic}{2}(\phi_2 - \phi_1) + \frac{\phi_2}{2}\left(1 - \frac{|\phi_2|^2}{4} + \eta \frac{|\phi_2|^4}{8}\right) = 0$$

(7)

Equations (7) are then simplified through standard polar transformations:

$$\phi_1 = R_1 e^{i\mu_1}, \quad \phi_2 = R_2 e^{i\mu_2}, \quad \gamma = \mu_1 - \mu_2 \tag{8}$$

Here $R_1, R_2, \mu_1, \mu_2$ are real functions. Substitution of (8) into (5) and rearrangement leads to the following set of equations:

$$\dot{R}_1 = -\frac{c}{2}R_2 \sin\gamma - \frac{R_1}{2}\left(1 - \frac{R_1^2}{4} + \eta\frac{R_1^4}{8}\right)$$

$$\dot{R}_2 = \frac{c}{2}R_1 \sin\gamma - \frac{R_2}{2}\left(1 - \frac{R_2^2}{4} + \eta\frac{R_2^4}{8}\right) \quad (9)$$

$$\dot{\gamma} = -\frac{c}{2}\cos\gamma \frac{\left(R_2^2 - R_1^2\right)}{R_1 R_2}$$

The inherent problem in representation (9) is the singularity of $\dot{\gamma}$ when $R_1 = 0$ or $R_2 = 0$. A typical behavior of $R_1, R_2, \gamma$ in the regime of non-conventional synchronization is presented in Figure 2. One can observe that the phase difference $\gamma$ undergoes periodic jumps from $\pi/2$ to $-\pi/2$ when $R_1$ or $R_2$ approach 0, just due to this singularity.

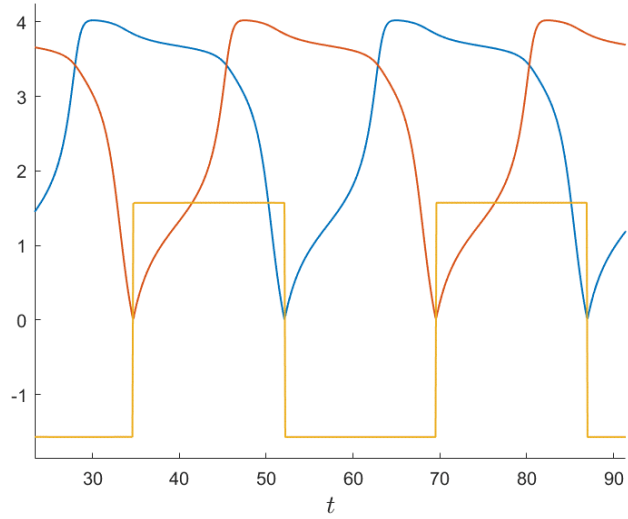

*Figure 2 – Slow flow of the strongly modulated response (model (9)), $R_1$ - blue, $R_2$ - red, $\gamma$ - yellow, Parameters: $c = 0.3, \eta = 0.1$.*

### 3. Analysis in modal coordinates

As an alternative approach to analysis of System (1), we suggest a transformation to modal coordinates.

$$\begin{aligned} u &= y_1 + y_2 \\ v &= y_1 - y_2 \end{aligned} \quad (10)$$

Here $u(t), v(t)$ represent the symmetric and the antisymmetric modes of linear counterpart in System (1), respectively. The coefficients in (10) are somewhat uncommonly set to unity and the transformation is not area-preserving, but this particularity is immaterial for the following analysis, since we look for the limit cycles and fixed points of the slow flow. In Figure 3 the response regime presented in Figure 1 is re-plotted in the modal coordinates in accordance with (10). One observes that the transition into modal space leads to a major simplification with respect to the physical variables. Namely, the regime corresponds to synchronization of the modulation amplitudes for both modes. In addition, it is easy to see that the relative phase between the modal coordinates is not constant and exhibits the drift. We note in passing that for the synchronization regimes with constant amplitude, the slow-flow modal coordinates will exhibit steady-state patterns with constant amplitudes and constant relative phase.

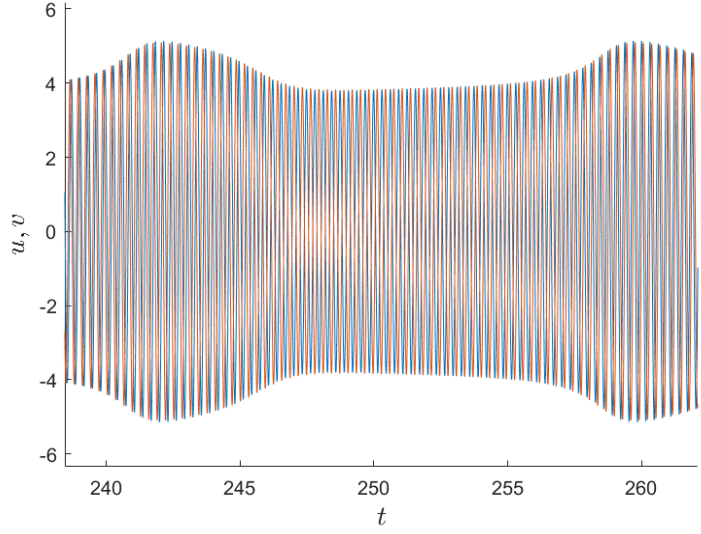

*Figure 3 – Strongly modulated response in modal space, u - blue, v - red (time rescaled by 20), Parameters: $c = 0.3$, $\eta = 0.1$, $\varepsilon = 0.05$.*

To explore the regime of the synchronization between the modal modulation amplitudes, we perform the averaging procedure for System (1) in the modal space and introduce new complex variables $\Psi_1, \Psi_2$:

$$u = -i\frac{\Psi_1 - \Psi_1^*}{2}, \quad u' = \frac{\Psi_1 + \Psi_1^*}{2}, \quad u'' = \Psi_1' - i\frac{\Psi_1 + \Psi_1^*}{2}$$
$$v = -i\frac{\Psi_2 - \Psi_2^*}{2}, \quad v' = \frac{\Psi_2 + \Psi_2^*}{2}, \quad v'' = \Psi_2' - i\frac{\Psi_2 + \Psi_2^*}{2} \tag{11}$$

By taking $\Psi_1 = \varphi_1 e^{it}$, $\Psi_2 = \varphi_2 e^{it}$, substitution of (10), (11) into (1) and averaging out the fast oscillation, the following set of equations for the slow flow of the modal coordinates is obtained:

$$\dot{\varphi}_1 + \frac{1}{2}\varphi_1 + \frac{1}{256}\left(-8\varphi_1|\varphi_1|^2 - 16\varphi_1|\varphi_2|^2 - 8\varphi_2^2\varphi_1^*\right) +$$
$$+ \frac{\eta}{256}\left(\varphi_1^3\varphi_2^{*2} + \varphi_1|\varphi_1|^4 + 3\varphi_1|\varphi_2|^4 + 2\varphi_2^3\varphi_1^*\varphi_2^* + 3\varphi_2^2|\varphi_1|^2\varphi_1^* + 6|\varphi_1|^2|\varphi_2|^2\varphi_1\right) = 0$$
$$\dot{\varphi}_2 - ic\varphi_2 + \frac{1}{2}\varphi_2 + \frac{1}{256}\left(-8\varphi_1^2\varphi_2^* - 16|\varphi_1|^2\varphi_2 - 8\varphi_2|\varphi_2|^2\right) + \tag{12}$$
$$+ \frac{\eta}{256}\left(\varphi_2^3\varphi_1^{*2} + \varphi_2|\varphi_2|^4 + 3\varphi_2|\varphi_1|^4 + 2\varphi_1^3\varphi_2^*\varphi_1^* + 3\varphi_1^2|\varphi_2|^2\varphi_2^* + 6|\varphi_1|^2|\varphi_2|^2\varphi_2\right) = 0$$

Then, we use transformation $\varphi_1 = N_1 e^{i\delta_1}$, $\varphi_2 = N_2 e^{i\delta_2}$, $\Delta = \delta_1 - \delta_2$ and obtain:

$$\dot{z}_1 + z_1 - \frac{z_1}{16}\left(z_1 + 2z_2 + z_2\cos(2\Delta)\right) + \frac{\eta z_1}{128}\left(\left(2z_2^2 + 4z_1 z_2\right)\cos(2\Delta) + z_1^2 + 3z_2^2 + 6z_1 z_2\right) = 0$$

$$\dot{z}_2 + z_2 - \frac{z_2}{16}\left(z_2 + 2z_1 + z_1\cos(2\Delta)\right) + \frac{\eta z_2}{128}\left(\left(2z_1^2 + 4z_1 z_2\right)\cos(2\Delta) + z_2^2 + 3z_1^2 + 6z_1 z_2\right) = 0 \quad (13)$$

$$\dot{\Delta} + c + \sin(2\Delta)\left[\frac{1}{32}(z_1 + z_2) - \frac{\eta}{128}(z_1 + z_2)^2\right] = 0$$

Here $z_1 = N_1^2$, $z_2 = N_2^2$. The structure of system (13) is symmetrical with respect to $z_1$ and $z_2$. Therefore, the amplitude synchronization ansatz $z_1 = z_2$ always satisfies this system of equations. These equations of the slow flow in modal space do not involve any singularities, contrary to the slow flow in physical coordinates (9). In Figure 4 we re-plot the regime with complete energy exchange in terms of the slow flow in the modal space (13). One observes the synchronization between $z_1, z_2$ and the smooth behavior of $\Delta$.

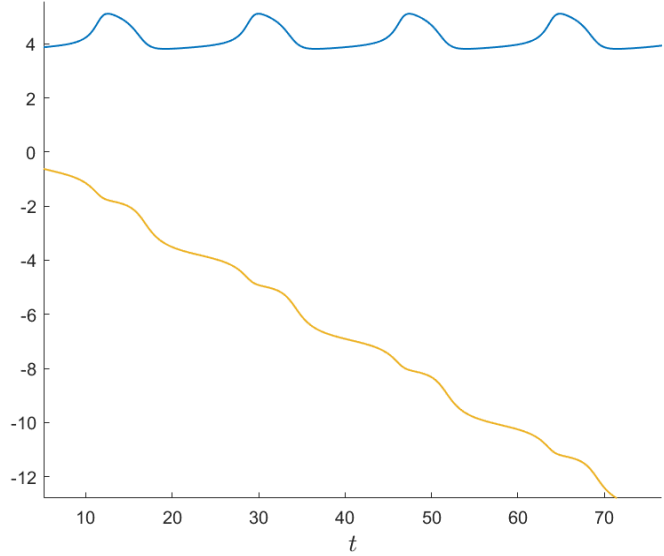

*Figure 4 – Slow flow of the regime with energy exchange in modal space, $z_1, z_2$ - blue, $\Delta$ - yellow, Parameters: $c = 0.3$, $\eta = 0.1$.*

Interestingly, fixed points of System (13) can be analyzed in exhaustive way despite high powers of the involved polynomials. From equations (13) one obtains the following algebraic conditions for the fixed points:

$$z_{1,0} - \frac{z_{1,0}}{16}\left(z_{1,0} + 2z_{2,0} + z_{2,0}\cos(2\Delta_0)\right) + \frac{\eta z_{1,0}}{128}\left(\left(2z_{2,0}^2 + 4z_{1,0} z_{2,0}\right)\cos(2\Delta_0) + z_{1,0}^2 + 3z_{2,0}^2 + 6z_{1,0} z_{2,0}\right) = 0$$

$$z_{2,0} - \frac{z_{2,0}}{16}\left(z_{2,0} + 2z_{1,0} + z_{1,0}\cos(2\Delta_0)\right) + \frac{\eta z_{2,0}}{128}\left(\left(2z_{1,0}^2 + 4z_{1,0} z_{2,0}\right)\cos(2\Delta_0) + z_{2,0}^2 + 3z_{1,0}^2 + 6z_{1,0} z_{2,0}\right) = 0$$

$$c + \sin(2\Delta_0)\left[\frac{1}{32}(z_{1,0} + z_{2,0}) - \frac{\eta}{128}(z_{1,0} + z_{2,0})^2\right] = 0$$

(14)

Cancelling $z_{1,0}, z_{2,0}$ from the first and the second equations respectively, and subtracting the remaining equations, one obtains:

$$\frac{1}{16}(z_{1,0} - z_{2,0})\left[\frac{\eta}{4}(z_{1,0} + z_{2,0}) - 1\right](1 + \cos(2\Delta_0)) = 0 \tag{15}$$

It is easy to see that the options $\cos(2\Delta_0) = -1$ or $z_{1,0} + z_{2,0} = 4/\eta$ are inconsistent with the third equation of System (14). Then, for the fixed points only three combinations of solutions are possible: 1) $z_{1,0} = z_{2,0}$, 2) $z_{1,0} = 0$, 3) $z_{2,0} = 0$. For the modes $z_{1,0} = 0$ or $z_{2,0} = 0$ we observed only trivial dynamics of localization on symmetric or antisymmetric mode; these regimes will stay beyond our scope. Moreover, only the dynamics of amplitude synchronization $z_1 = z_2 \equiv z$ will be considered. The stable solutions that will be obtained under this assumption need to be validated in the original modal system (1) for global stability. Dynamics of the amplitude synchronization is therefore described by the following reduced model:

$$\begin{aligned}
&\dot{z} + z - \frac{z^2}{16}(3 + \cos(2\Delta)) + \frac{\eta z^3}{64}(5 + 3\cos(2\Delta)) = 0 \\
&\dot{\Delta} + c + \frac{\sin(2\Delta)}{32}(2z - \eta z^2) = 0
\end{aligned} \tag{16}$$

This system defines the phase flow on cylinder $\{0 \leq \Delta < \pi, z > 0\}$ and has two control parameters ($c, \eta$). The map of regimes at the plane of these control parameters is presented in Figure 4.

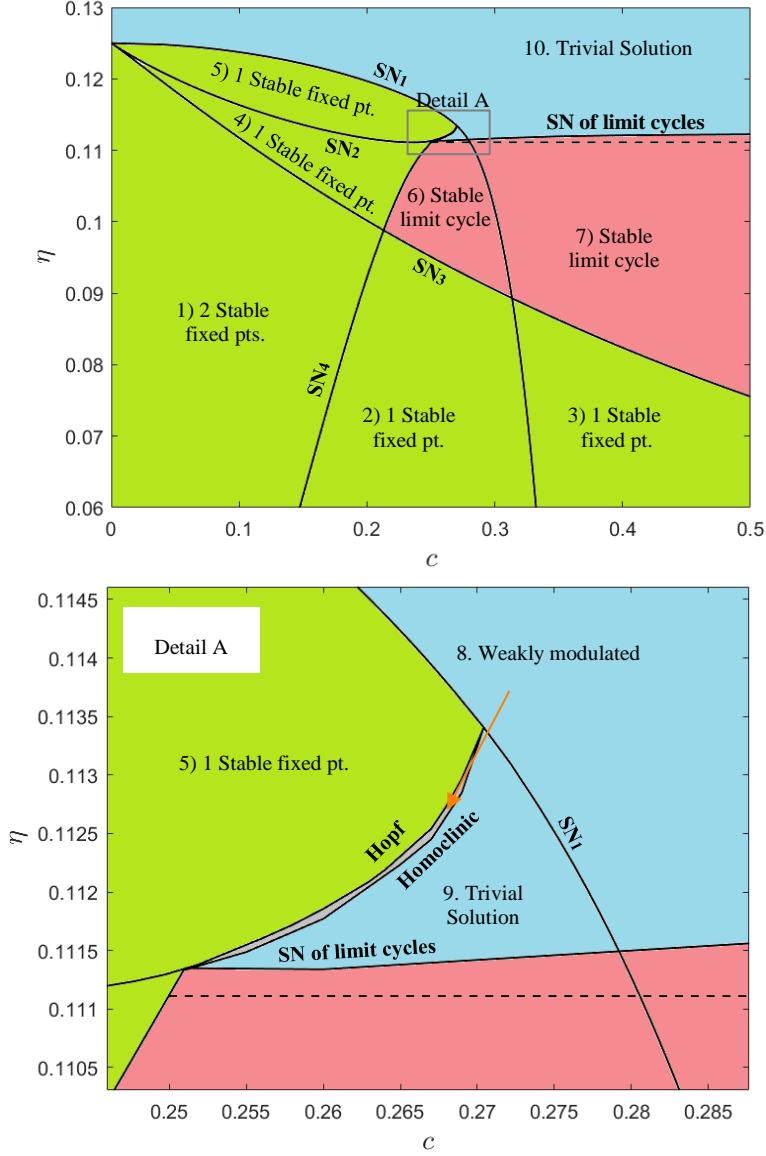

*Figure 5 – Bifurcations and solution zones of system (16) for the case $z_1 = z_2$*

In Figure 5 various regimes and their boundaries are described. First, we address non-trivial fixed points of System (16), described by the algebraic equations:

$$z_0 - \frac{z_0^2}{16}\left(3+\cos(2\Delta_0)\right) + \frac{\eta z_0^3}{64}\left(5+3\cos(2\Delta_0)\right) = 0$$
$$c + \frac{\sin(2\Delta_0)}{32}\left[2z_0 - \eta z_0^2\right] = 0 \qquad (17)$$

These fixed points correspond to common synchronized limit cycle oscillations (LCO) with constant amplitudes in the original system, since the phase difference between the modes is also stationary. Simple algebraic manipulations yield:

$$\left(\frac{32c}{\eta z_0^2 - 2z_0}\right)^2 + \left(\frac{5\eta z_0^2 - 12z_0 + 64}{4z_0 - 3\eta z_0^2}\right)^2 = 1$$

$$\Delta_0 = \frac{1}{2}\sin^{-1}\frac{32c}{\eta z_0^2 - 2z_0} \qquad (18)$$

Let us examine the following polynomial obtained from (18):

$$P_1(z_0) = \left(\eta z_0^2 - 2z_0\right)^2 \left[\left(5\eta z_0^2 - 12z_0 + 64\right)^2 - \left(4z_0 - 3\eta z_0^2\right)^2\right] + 32^2 c^2 \left(4z_0 - 3\eta z_0^2\right)^2 \qquad (19)$$

By taking $P_1(z_0) = 0, \frac{dP_1(z_0)}{dz_0} = 0$ one obtains four different saddle-node (SN) limits, at which two fixed points collide and disappear. The SN limits are plotted in Figure 5.

The next issue to address is the stability of the fixed points. For this sake, we rewrite system (13) in the following way:

$$\dot{z}_1 = -z_1 + \frac{z_1}{16}\left(z_1 + 2z_2 + z_2 \cos(2\Delta)\right) - \frac{\eta z_1}{128}\left(\left(2z_2^2 + 4z_1 z_2\right)\cos(2\Delta) + z_1^2 + 3z_2^2 + 6z_1 z_2\right) \equiv f_1(z_1, z_2, \Delta)$$

$$\dot{z}_2 = -z_2 + \frac{z_2}{16}\left(z_2 + 2z_1 + z_1 \cos(2\Delta)\right) - \frac{\eta z_2}{128}\left(\left(2z_1^2 + 4z_1 z_2\right)\cos(2\Delta) + z_2^2 + 3z_1^2 + 6z_1 z_2\right) \equiv f_2(z_1, z_2, \Delta)$$

$$\dot{\Delta} = -c - \sin(2\Delta)\left[\frac{1}{32}(z_1 + z_2) - \frac{\eta}{128}(z_1 + z_2)^2\right] \equiv f_3(z_1, z_2, \Delta)$$

$$(20)$$

The Jacobian for the explored case of $z = z_1 = z_2$ at the fixed points is given by:

$$J_0 = J\big|_{z_1 = z_0, z_2 = z_0, \Delta = \Delta_0} = \begin{bmatrix} \frac{\partial f_1}{\partial z_1} & \frac{\partial f_1}{\partial z_2} & \frac{\partial f_1}{\partial \Delta} \\ \frac{\partial f_2}{\partial z_1} & \frac{\partial f_2}{\partial z_2} & \frac{\partial f_2}{\partial \Delta} \\ \frac{\partial f_3}{\partial z_1} & \frac{\partial f_3}{\partial z_2} & \frac{\partial f_3}{\partial \Delta} \end{bmatrix}_{z_1 = z_0, z_2 = z_0, \Delta = \Delta_0} \qquad (21)$$

The eigenvalues for the Hopf bifurcation are purely imaginary $\lambda = \pm i\omega$. Therefore, the following two equations can be written at this limit:

$$\text{Re}|J_0 - i\omega I| = 0$$
$$\text{Im}|J_0 - i\omega I| = 0 \qquad (22)$$

We solve numerically the set of 4 equations (22), (18) and get the boundary of the Hopf bifurcation illustrated in Figure 5.

Regimes of amplitude synchronization with phase drift correspond to limit cycles of System (16). Due to the cylindrical structure of the phase space, one can encounter a LCO of the first type (not winding the phase cylinder) and of the second type (winding the cylinder once) [24, 25]. These LCO are created through bifurcational mechanisms depicted in Figure 5. Namely, the LCO of the first type appears through Hopf bifurcation and disappears through the homoclinic global bifurcation. LCO of the second type appears due to the saddle-node bifurcation of limit cycles.

The dashed line in Figure 5 at $\eta = 1/9$ and $c > 1/4$ corresponds to the unique case of the special symmetry explored in [15], when the LCO can be described at the phase plane even in physical coordinates. On this line the value of z is constant and equal to 12 while $\Delta$ is time dependent. The solution for this case is given in (23). Interestingly, despite the additional symmetry, no bifurcation is encountered at this line besides the terminal point c=1/4. However, we observe in Figure 5 that the SN bifurcation of the LCO of the second type occurs very close to this line. This proximity requires further analysis and understanding.

$$z = 12, \quad \Delta = -\tan^{-1}\left(\frac{\tan\left(\frac{(t-t_0)}{4}\sqrt{16c^2-1}\right)\sqrt{16c^2-1}+1}{4c}\right) \qquad (23)$$

One can see that the system exhibits a rich set of responses and bifurcations. The parametric $c - \eta$ space in Figure 5 is split into 10 different regions with the following characteristics:

- Nontrivial stable fixed points – regions 1-5 (shaded with green). In these regions one or two fixed points are stable alongside the always stable trivial solution.
- Limit cycle of the second type – regions 6-7 (shaded with red). In these regions a limit cycle is stable alongside with the trivial solution.
- Limit cycle of the first kind – region 8 (shaded with gray). This narrow region is bounded by lines of the Hopf bifurcation and of the global homoclinic bifurcation. This limit cycle is also stable alongside with the trivial solution.
- Trivial solution – regions 9-10 (shaded with blue).

This list represents the possible regimes of amplitude synchronization in the modal space of the system. Successive transitions from region 5 to region 9 are illustrated in Figure 6. In the current representative example, parameter $c = 0.26$ is kept constant and parameter $\eta$ decreases. In region 5 one encounters a stable focus and a saddle. Transition into region 8 occurs through a Hopf bifurcation, at which the stable focus becomes unstable and a limit cycle of the first kind, marked with green, is formed. Then, a global homoclinic bifurcation is reached and the limit cycle merges with the stable and unstable manifolds of the saddle and disappears. Further decrease of $\eta$ leads into region 9 where one observes an unstable focus and a saddle. All initial conditions on the cylinder lead to the lone stable attractor of the amplitude death.

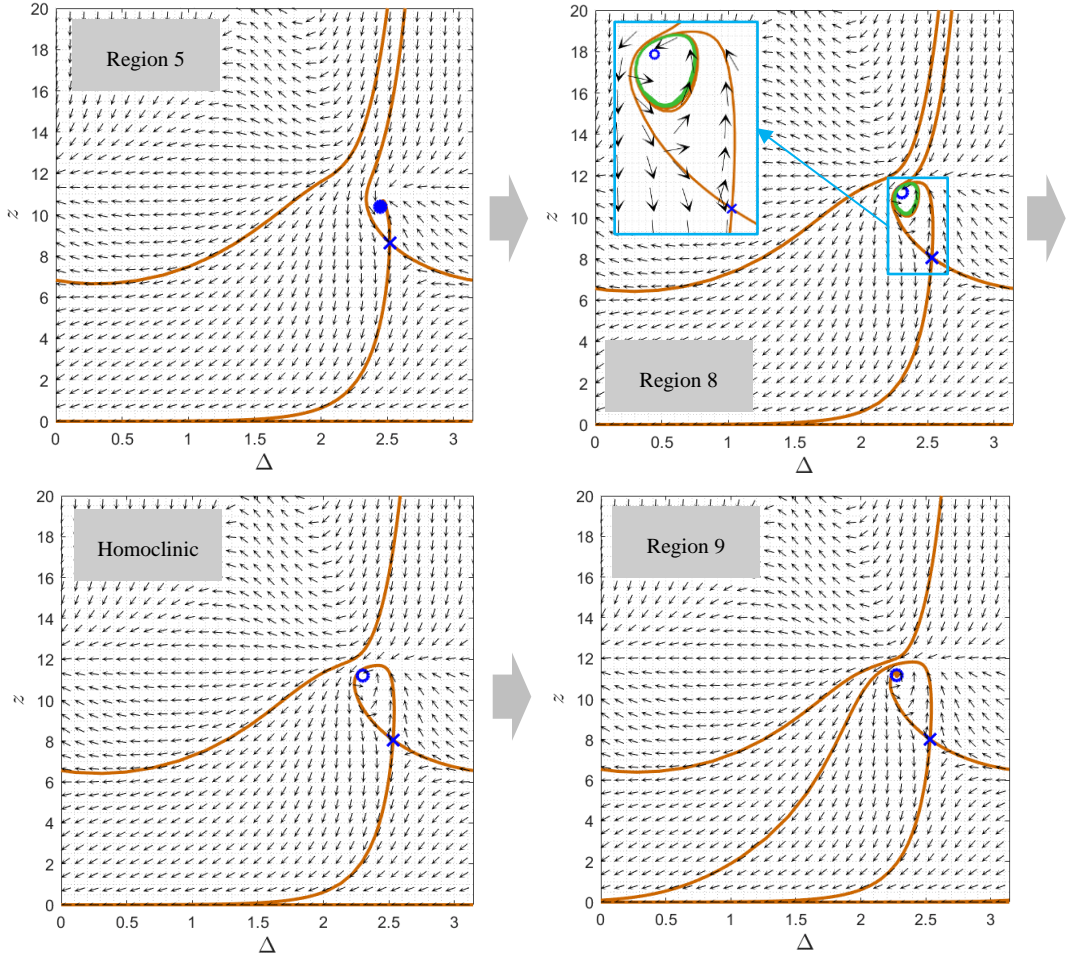

*Figure 6 – Transition from region 5 into region 9 through Hopf and homoclinic bifurcations. Parameters: Region 5: $c = 0.26, \eta = 0.114$, Region 8: $c = 0.26, \eta = 0.1118$, Homoclinic: $c = 0.26, \eta = 11177191$, Region 9: $c = 0.26, \eta = 0.1115$*

Then, we describe the creation of the limit cycle of the second kind. At the upper limit, the limit cycle is created through a SN bifurcation of limit cycles in transition between regions 9 and 10 to regions 6 and 7 respectively. The transition from region 10 to 7 is illustrated in Figure 7.

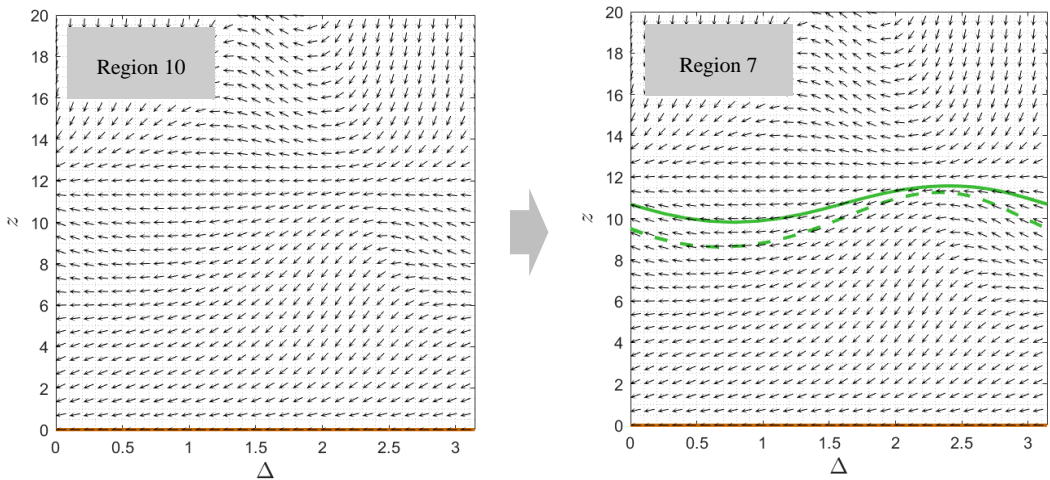

*Figure 7 – Transition from region 10 into region 7 through a saddle-node of limit cycles bifurcation. Parameters: Region 10: $c = 0.4$, $\eta = 0.113$, Region 7: $c = 0.4$, $\eta = 0.112$.*

On the other limits, from the left and from below, the limit cycle of the second kind (marked with green) is created through SN bifurcation – the saddle and the node disappear and form the cycle. This case is demonstrated in Figure 8.

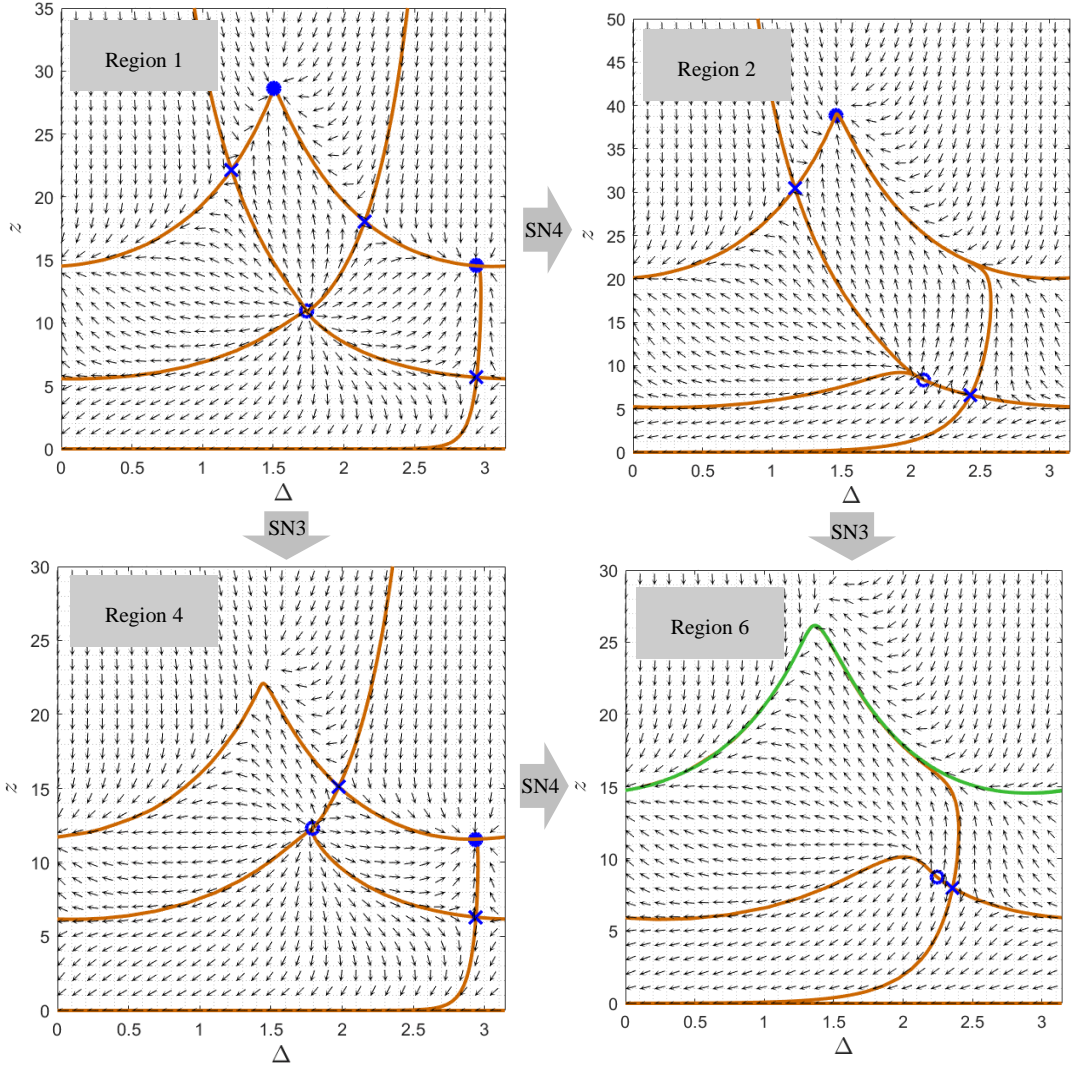

*Figure 8 – Transition from region 1 into region 6 through saddle-node bifurcations on the limit cycle. Parameters: Region 1: $c=0.1, \eta=0.1$, Region 2: $c=0.3, \eta=0.08$, Region 4: $c=0.1, \eta=0.113$, Region 6: $c=0.3, \eta=0.1$.*

## 4. Numerical verification of the results

So far, the qualitative analysis of system (16) seems complete. In order to reach system (16), two major simplifications were conducted: averaging and the reduction to the special case of amplitude synchronization in modal space $z=z_1=z_2$. The next step is to validate these assumptions, by examining the stability of the unique attractors of the system in the original system ((1) and (10)) and in the full averaged system (13).

For this goal we perform numerical tests in the interesting section $c=0.26$ where most of attractors can be found within minor variations of parameter $\eta$. The first attractor to examine is the limit cycle of the second kind response that is expected according to the analysis, for instance, at $\eta=0.1112$. This case is shown for the original coordinates $y_1, y_2$ (according to Equation (1)) in Figure 9a and in modal space

(10) for coordinate $u$ alongside the envelope obtained from the full averaged model (13) in Figure 9b. The averaged model follows exactly the envelope of the response. We again observe that extreme energy exchanges between the physical oscillators correspond to only relatively weak modulation in the modal space. The second attractor, expected for a representative value $\eta = 0.11178$, is a limit cycle of the first kind, contained in a narrow region between the Hopf and homoclinic bifurcations in Figure 5. As expected, we find such a response (Figure 10) for both the modal system and the full averaged system. Yet again a perfect match is obtained between the averaged and the original modal system. Generally, the provided numerical verifications imply that the analysis of the simplified system (16) as outlined in Figure 5 faithfully represents the features of the original system.

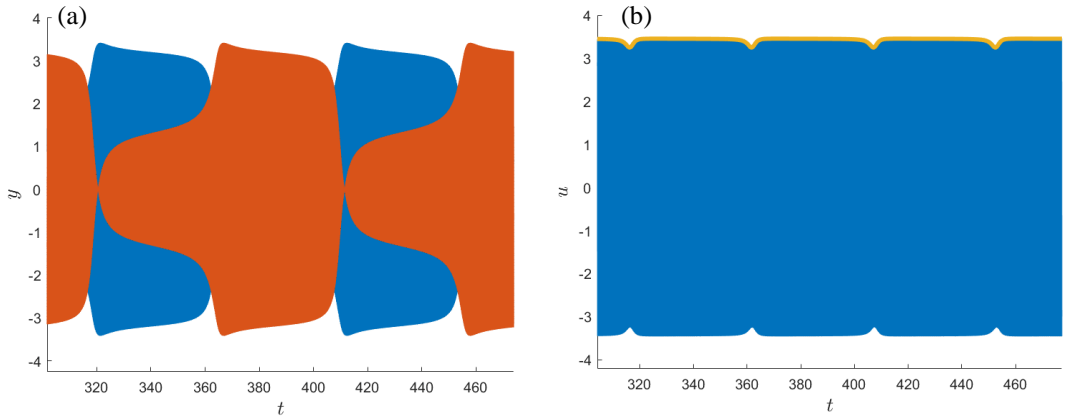

*Figure 9 – Limit cycle of the second kind of the slow flow, (a) Original coordinates, $y_1$ - blue, $y_2$ - red (time recalled by 1000), (b) Modal coordinates, full response – blue (time recalled by 1000), averaged envelope (13) – orange, Parameters: $c = 0.26, \eta = 0.1112, \varepsilon = 0.001$.*

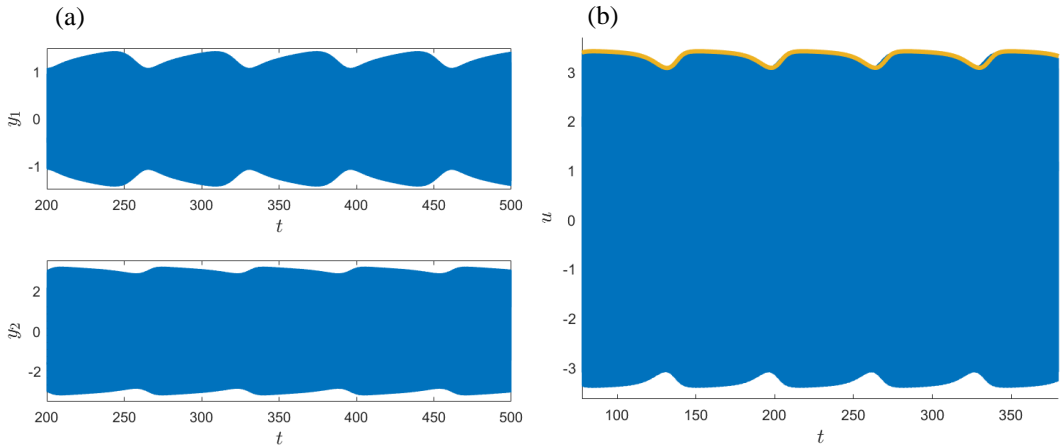

*Figure 10 – Limit cycle of the first kind of the slow flow, (a) Original coordinates (time recalled by 1000), (b) Modal coordinates, full response – blue (time recalled by 1000), averaged envelope (13) – orange, Parameters: $c = 0.26, \eta = 0.11178, \varepsilon = 0.001$.*

The results in Figure 9 for the LC of the second kind are similar to large extent to the situation described in Section 2. The results for the LC of the first kind in Figure 10 are somewhat surprising. One can observe that weakly modulated amplitude-synchronized LCO in the modal space (Figure 10b) correspond to *asymmetric* torus in the physical space. Equations (1) are symmetric, and therefore substitution $y_1 \leftrightarrow y_2$ will yield *different* solution in the physical space with *the same* slow-flow envelope in the modal space. Mathematically it is possible since the transformation from 4-dimensional state space of the original system to 3-dimensional slow-flow equations in injective, but obviously not bijective.

## 5. Concluding remarks.

The results presented above reveal the underlying mechanism of the "non-conventional" synchronization in a system of weakly coupled bi-stable van der Pol oscillators. This regime, in fact is a LCO in the slow-flow space with synchronization of the modulation amplitudes between the symmetric and antisymmetric modes of the system. Averaging in the modal space substantially simplifies the treatment and removes the singularities from the slow-flow equations. Moreover, these slow-flow equations in the modal space possess sufficient internal symmetry to reduce the dynamics to a simple phase cylinder for any set of the control parameters. Therefore, it is possible to investigate the structure of the parametric space and to explore the set of response scenarios and their local and global bifurcations. Previously observed "nonconventional synchronization" with almost complete energy exchange between the oscillators correspond to the LCO of the second kind on the phase cylinder. In addition, one can observe the regime of asymmetric nonstationary synchronization in the physical space depicted in Figure 10, that corresponds to the LC of the first kind for the slow flow in the modal space. To the best of the author's knowledge, this latter regime has not been observed in previous studies on the subject.

One can expect that the observed regimes will appear generically in similar systems of coupled oscillators with self-excitation. As for the considered system, the unresolved issues include the role of special analytically derivable limit cycle for $\eta = 1/9$. Both global bifurcations (homoclinic and SN of the LCO) occur not on this line, but in its close vicinity. Unusual sensitivity of responses on miniscule variation of the control parameters allows one to expect that further simplifications in this and similar systems are possible. This issue will be the subject of further investigations.

**Acknowledgment**

The authors are very grateful to Israel Science Foundation (grant 1697/17) for financial support.

**Conflict of interests**

The authors declare no conflict of interests.